\documentstyle[epsfig]{aipproc}
\newcommand{\msun}[0]{M_\odot}
\begin{document}
\title{LIGO's ``Science Reach''\thanks{With permission from
    ``Astrophysical Sources of Gravitational Radiation for
    Ground-Based Detectors,'' copyright American Institute of Physics,
    2001.}
}
\author{Lee Samuel Finn}
\address{$^*$Center for Gravitational Physics and Geometry,\thanks{Also 
Department of Physics, Department of Astronomy and Astrophysics} The 
Pennsylvania State University, University Park, Pennsylvania 
16803}

\maketitle

\begin{abstract}
Technical discussions of the Laser Interferometer Gravitational Wave
Observatory (LIGO) sensitivity often focus on its 
effective sensitivity to gravitational waves in a given band;
nevertheless, the goal of the LIGO Project 
is to ``do science.''  Exploiting this new
observational perspective to explore the Universe is a long-term goal,
toward which LIGO's initial instrumentation is but a first step. 
Nevertheless, the first generation LIGO instrumentation is sensitive
enough that even non-detection --- in the form of an upper limit ---
is also informative.  In this brief article I describe in quantitative
terms some of the science we can hope to do with first and future
generation LIGO instrumentation: it short, the ``science reach'' of
the detector we are building and the ones we hope to build.
\end{abstract}

\section*{Introduction}

Technical discussions of the Laser Interferometer Gravitational Wave
Observatory (LIGO) sensitivity often focus on its effective
sensitivity to gravitational waves in a given band; nevertheless, the
LIGO Projects goal is to enable a new kind of astronomy ---
gravitational wave astronomy --- and explore the fundamental physics
of gravity.  The gravitational waves accessible to LIGO are of
astronomical origin, arise in regimes of strong, dynamical gravity,
and carry with them an imprint of their origin and gravity's character
in that regime.  From detection we aim to learn about the fundamental
physics of gravity and the astronomical character of the sources, and
we build LIGO in service of that goal.

Exploiting this new observational perspective to explore the Universe
is a long-term goal, toward which LIGO's initial instrumentation is
but a first step. To fully develop this new observational perspective
will require detectors of increasing sensitivity and detectors that
explore different spectral bands.  Nevertheless, the first generation
LIGO instrumentation is sensitive enough that even non-detection ---
in the form of an upper limit --- is also informative.  In this brief
article I describe in quantitative terms some of the science we can
hope to do with first and future generation LIGO instrumentation: it
short, the ``science reach'' of the detector we are building and the
ones we hope to build.

Strong gravitational wave sources have characteristics that make them
weak electromagnetic observations, and \emph{vice versa.} An immediate
consequence is that we generally know very little about the most
intriguing gravitational wave sources.  In choosing the source science
that I discuss below, I've deliberately taken a conservative approach
and focused my attention where speculation can be minimized.  I've
also chosen to stay ``close to the detector'': i.e., to focus on the
science that comes from direct detection or upper limits, as opposed
to interpretation of observations in terms of detailed
astrophysical models.

Still, we must remember that ``[i]n the field of observation, chance
favors the prepared mind'' \cite{pasteur1854a}: historically, the
opening of a new spectrum to observation has lead to the revision of
theoretical models, the development of new ideas, and, frequently,
major serendipitous discoveries.  LIGO is also a reach into the
unknown and it is there the most exciting prospects lay.

A familiarity with the anticipated noise character of the initial and
advanced LIGO designs is critical to an appreciation of their
 sensitivity to different sources of gravitational radiation;
for that background I refer the reader to Peter Fritschel's
contribution to this volume.  In the remainder of this article I
discuss first compact binary inspiral, focusing on the effective
volume of space that LIGO can survey for these sources, second on the
limits that LIGO can place the crustal deformation of young pulsars, 
thirdly on the energy density in a stochastic
gravitational wave background, fourthly on the limits that can be placed
on the gravitational radiation that might accompany $\gamma$-ray
bursts, and lastly on the limits that can be placed on the energy
radiated in a stellar core-collapse supernova.


\section*{Compact binary inspiral}

Most LIGO stories begin with compact binary inspiral.  Binary systems
consisting of neutron stars or stellar mass black holes decay owing to
the emission of gravitational radiation.  During the last few moments
in the life of these binary systems the gravitational radiation,
emitted at twice the orbital frequency, races through the LIGO band before
the binary components collide and coalesce into a final black hole
(or, perhaps, another, more massive neutron star supported against
collapse by its angular momentum).  The detailed character of the
radiation emitted during this \textit{inspiral\/} phase is generally
accessible to theoretical calculation and, correspondingly, has been
studied in great detail (cf.\ \cite{blanchet95a,blanchet98a} and
references therein). 

As inspiral proceeds the binary system becomes more compact and,
correspondingly, more relativistic and less accessible to a
perturbative, post-Newtonian treatment.  Ultimately, there is a ``last
orbit'' before the two components plunge toward each other and
coalesce.  Whether the transition from inspiral to plunge is due to a
dynamical instability, a secular instability, or simply dissipative
forces (e.g., radiation reaction) is an open question, as is the
detailed character of the radiation during plunge and coalescence. 

For black holes, coalescence represents the ultimate expression of
dynamical, strong field gravity.  It is certainly also accompanied by
a burst of radiation \cite{flanagan98a,flanagan98b}; however, that
burst is extremely difficult to model: the spacetime is dynamical, the
fields are strong and the structure of the binary components plays an
important role.  Modeling \emph{coalescence} (which we take to include
the plunge from inspiral) has been a long-term goal of numerical
relativity, and recent results suggest that the radiation emitted
during this phase alone may be as greater as 1\% or more of the
systems rest mass \cite{khanna99b}.

Lastly, the final remnant of the coalescence will be strongly
perturbed from equilibrium and will shed that perturbation, in part
through gravitational radiation.  This \textit{ring down} phase is, by
definition, perturbative and the waveform well understood (even if it
is not particularly well structured). 

The inspiral, coalescence and ring down are consecutive parts of the
signature of an inspiraling compact binary system.  A complete search
for compact binary inspiral will look for all three components and not
single out any given component, ignoring the rest.  Nevertheless, the
inspiral waveform is better understood, will appear in the LIGO
detectors at an earlier time, and carry more information about the
system, then the signals arising from coalescence or ring down.  In
the LIGO detectors the inspiral of a binary neutron star system
deposits most all of its contribution to the signal-to-noise as the
orbital frequency increases from approximately 20~s${}^{-1}$ to about
100~s${}^{-1}$ over approximately 20~s \cite{finn96a} immediately
preceding the coalescence\footnote{The band shifts slightly to lower
  frequencies and longer periods between the first and later
  generation LIGO detectors, corresponding to the greater
  low-frequency sensitivity of these more advanced detectors.};
additionally, the inspiral waveform carries information about the
system's component masses, redshift and luminosity distance
\cite{schutz86a,finn93a,cutler94a,finn96a}.

For the purpose of comparing the science reach of alternative
detectors, then, we will focus on just the radiation associated with
the inspiral of two $1.4\,\mbox{M}_{\odot}$ neutron stars or black
holes.  Assuming that we use matched filtering to detect the inspiral
radiation, and that we are contending only with Gaussian detector
noise, a false alarm rate of $10^{-4}\,\mbox{y}^{-1}$ corresponds to a
signal-to-noise threshold of approximately $8$ for detection.  With
this criteria we can calculate the observed rate of binary inspiral
$\dot{N}$ given the rate density (on a per co-moving time, co-moving
volume, basis) $\dot{n}$ of inspiraling binary systems and taking into
account the anisotropic radiation pattern of the sources and antenna
pattern of the detectors.  The ratio $\dot{N}/\dot{n}$ we
\textit{define} to be an \textit{effective} volume surveyed by the
detector, which we characterize by an effective radius,
$r_{\mbox{\small eff}}$:
\begin{equation}
    \dot{N} = {4\pi r^{3}_{\mbox{\small eff}}\over 3}\dot{n}.
\end{equation}

\begin{table}
    \caption{The effective volume of space surveyed for \emph{just the
      compact binary inspiral waveform} by the initial and
    --- proposed --- advanced LIGO detector system (sometimes known as
    LIGO II).  The volume is represented by an effective radius $r$,
    such that $4\pi r^{3}/3$ is the survey volume.  For both cases
    survey volumes are given for a single interferometer and for an
    effective interferometer synthesized by the phase coherent sum of
    the output of the three LIGO interferometers. For advanced LIGO 
    we assume that the 2~Km Hanford detector is upgraded to a 4~Km 
    detector.}\label{tbl:cbi}
    \begin{tabular}{lrrr}
        &NS/NS&NS/10 M$_{\odot}$ BH&10 M$_{\odot}$ BH/10 M$_{\odot}$BH\\
        \multicolumn{4}{l}{Initial LIGO}\\
        $\quad$Single IFO&14~Mpc&30~Mpc&72~Mpc\\
        $\quad$Three IFO&21~Mpc&44~Mpc&104~Mpc\\
        \multicolumn{4}{l}{Advanded LIGO}\\
        $\quad$Single IFO&200~Mpc&420~Mpc&620~Mpc\\
        $\quad$Three IFO&350~Mpc&740~Mpc&1.0~Gpc
    \end{tabular}
\end{table}

Table \ref{tbl:cbi} gives the effective radius $r_{\mbox{\small eff}}$
for the LIGO initial and (as of this writing) advanced detector
concepts.  Over the volume surveyed by initial LIGO only the most
optimistic scenarios give a reasonable event rate for NS/NS binary
inspiral; however, advanced LIGO should observe many inspirals, of
different types, per year (see Kalogera, this proceedings).  An
interesting note is that, in terms of the expected binary inspiral
detection rate, it is more likely that a NS/NS inspiral will be
detected in 3~h of advanced LIGO operations than in 1~y of initial
LIGO operations! 

\section*{Periodic Signals}

Shortly after their formation, rapidly rotating neutron stars develop
a solid crust.  As they evolve, they gain angular momentum through
accretion, lose it through radiation, and see it redistributed between
the fluid interior and the solid crust.  Through these processes, as
well as crust fracturing and the neutron star equivalent of
``continental drift'' \cite{ruderman91a,ruderman91b,ruderman91c},
neutron stars cease being axisymmetric and will radiate periodic
gravitational waves with an amplitude proportional to the crust
strain.  The amplitude of the gravitational waves depends on the stars
period and the degree of non-axisymmetry.  If the star is rotating
about a principal axis of its moment of inertia the radiation period
will be twice the rotational period and its amplitude proportional to
the the difference between the moment of inertia of the other two
principal axes.

Theoretical prejudice regarding the maximum sustainable crust stress
bound this difference to less than or of order $10^{40}\,\text{g
  cm}^2$: {\em i.e,} something less than one part in $10^{5}$ of the
moment of inertia itself \cite{alpar85a}.\footnote{Observations of the
period, spin-down and rate of spin-down of very old pulsars places
limits on $\epsilon\equiv|I_2-I_1|/I_3$ of less than or of order
$10^{-8}$ for these stars; however, owing to their great age and past
history the crust of these stars has been heavily annealed. Thus, this
limit is not likely representative of young neutron stars.}
Gravitational wave observations over an extended period will, in
principle, be able to bound the \emph{actual} strain in nearby pulsars
to an order of magnitude greater than this theoretical bound.

\begin{figure}
\epsfxsize=\columnwidth
\begin{center}
\epsffile{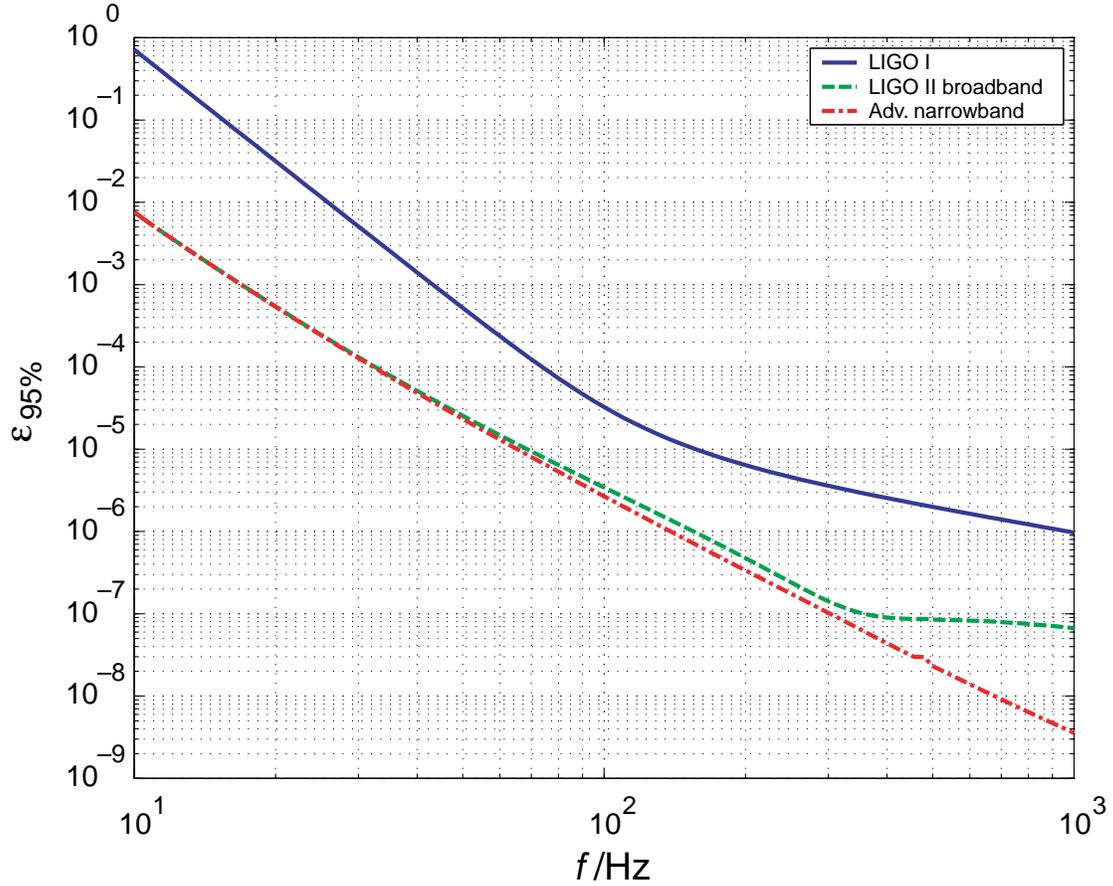}
\end{center}
\caption{The ``science reach'' of the initial and advanced LIGO
  detectors, when applied to the determination of a pulsar's triaxiality
  parameter $\epsilon$. Shown is the upper limit (95\% confidence)
  that can be set on $\epsilon$ in a 1~y observation of a pulsar at a
  distance of 10~Kpc, as a function of the gravitational wave
  frequency $f$. The solid curve is the upper limit that can be set by
  the initial LIGO detectors; the dashed curve the limit that can be
  set by the proposed advanced LIGO detectors, operating in a
  broadband mode, and the dot-dashed curve  the limit that can be
  set by the proposed advanced LIGO detector if its sensitivity at any
  given frequency were tuned so that the photon shot noise were
  negligible.}\label{fig:pulsar}
\end{figure}

To describe the LIGO detector's science reach in the context of
periodic signals, we ask what limit can be placed on triaxiality
parameter $\epsilon$ ($|I_2-I_1|/I_3$) for a pulsar, rotating about
the principal axis with moment of $I_3$. This limit will depend on the
pulsar period, detector noise at the gravitational wave period (half
the pulsar period), the distance to the pulsar, and the pulsar's
declination. For the purpose of assessing the science reach, focus
attention on a source at a distance of 10~Kpc and average over the
source declination. Figure \ref{fig:pulsar} shows the science reach of
the initial and advanced LIGO detectors. Three curves are shown: the
solid curve shows the reach of the initial LIGO detector, which is
clearly in a position to place interesting, but not challenging, upper
limits on pulsars with frequencies $\sim100$~Hz. The dashed curve
shows the reach of the advanced LIGO detector, operating in a
broadband mode tuned to maximize the detector's reach for binary
inspiral (as described in the previous section). 

Finally, the dash-dot curve shows the science reach for an advanced
detector whose optical configuration has been tuned to maximize the
its sensitivity at the indicated frequency. Here we go beyond what is
proposed for the next generation LIGO detectors. That detector will be
tunable; however, the ability to reach the limit shown depends on
choices made in the mirror coatings, which are different if one
optimizes for, say, 500~Hz to 1~KHz frequencies as opposed to 100~Hz
to 500~Hz, and also the quality of the optical components. The
dash-dot curve is the envelope of best possible behavior: i.e.,
limited at each frequency by the thermal noise of the proposed,
advanced detector. The actual proposed advanced detector, optimized
for detection at a particular frequency, will have a science reach
bounded from below by the dash-dot curve and above by the dashed
curve. Particularly for pulsars closer than the galactic center, the
advanced LIGO sensitivity gives us the ability to place challenging
upper limits on the crustal deformation in younger pulsars.

\section*{$\gamma$-ray bursts}

$\gamma$-ray bursts are sudden, intense flashes of $\gamma$-rays
lasting anywhere from a fraction of a second to hundreds of
seconds. The burst sources are 
confirmed through observations to be at cosmological
distances and involve power outputs of $10^{51}$--$10^{54}$~erg/s,
which is comparable to the total conversion of the Sun's rest-mass to
$\gamma$-rays over the course of a few seconds or to the emission over
the same period of as much energy as our entire galaxy radiates in
100~y. This is a far larger luminosity than that of any other known
astronomical source \cite{meszaros01a}.

The progenitors of $\gamma$-ray bursts are not yet well identified and
there is good reason to believe they may be a heterogeneous population
\cite{kouveliotu93a,lamb93a,mukherjee98a,meszaros00a}.  Two classes of
progenitors are currently favored: neutron star binary coalescence or
hypernovae or collapsars
\cite{woosley93a,paczynski86a,paczynski98a,fryer99a,goodman86a,meszaros97a}.
All models converge on the formation of a several solar mass black
hole, surrounded by a debris torus whose accretion provides the energy
necessary to power the $\gamma$-ray burst.  The $\gamma$-ray burst
itself arises as a result of either internal shocks \cite{rees94a}
(owing to velocity variations) in the outgoing fireball and/or the
formation of forward and reverse shocks when the expanding fireball
impacts on, e.g., the interstellar medium.

The violent formation of a black hole will almost certainly involve a
gravitational wave burst. The qualitative character of that
burst will likely depend on the progenitor: e.g., the spectrum and
timescale of the burst arising from a hypernova, or collapsar, or the
coalescence of a compact binary will all likely be different. The
interval between the gravitational and $\gamma$-ray bursts will depend
on whether the $\gamma$-ray burst is formed via internal shocks
(timescales on order $0.1$~s) or the formation of a blast wave and
reverse shock as the fireball impacts on an external medium (timescales
on order $100$~s).  Correspondingly, observations of correlated
gravitational and $\gamma$-ray bursts can
\emph{i)} test the general model of $\gamma$-ray burst sources (i.e.,
the formation of a several $\msun$ black hole),
\emph{ii)} potentially distinguish between different progenitors 
(e.g., hypernovae, collapsar or compact binary coalescence), 
\emph{iii)} determine whether the $\gamma$-rays are produced via
internal or external shocks (via the interval between the $\gamma$-ray
and gravitational wave burst).

Since GRBs occur at cosmological distances the signal-to-noise
ratio~(SNR) of any individual GWB will likely be insufficient for
direct detection in either the initial or advanced LIGO detectors.
Nevertheless, it will still be possible to detect a \emph{statistical
  association} between gravitational wave and $\gamma$-ray bursts.  If
GWBs are associated with GRBs, the correlated output of two GW
detectors will be different in the moments immediately preceding a GRB
(\emph{on-source}) than at other times not associated with a GRB
(\emph{off-source}). (While we focus on $\gamma$-ray bursts here, any
plausible class of astronomical events can serve as a trigger.) A
statistically significant difference between on- and off-source
cross-correlations would support a GWB/GRB association and represent a
detection of gravitational waves by the detector pair.  We can measure
this difference using Student's $t$-test without requiring any
foreknowledge of the signal waveform, source or source population
(though with such a model the effectiveness of the test can be
improved). The measured difference can be used to establish a
confidence interval (CI) or upper limit (UL) on the rms amplitude of
GWBs associated with GRBs. The CI/UL, in turn, constrains any model
for model for GRB/GWB pairs \cite{finn99f}. The 95\% confidence upper
limit $h_{95\%}$ we can 
set on the strength of the gravitational waves associated with
$\gamma$-ray bursts, is given by
\begin{equation}
h_{95\%}^2 = \left\{
\begin{array}{l}
\left(1.35\times10^{-22}\right)^2\\
\left(2.5\times10^{-23}\right)^2
\end{array}
\right\}
\left({T\over 0.2\,\mbox{s}}{1000\over N_{\mbox{\small on}}}\right)^{1/2}
\qquad
\begin{array}{l}
\mbox{initial LIGO}\\
\mbox{advanced LIGO}
\end{array}
\end{equation}
where $T$ is the delay between the gravitational wave and $\gamma$-ray
burst (0.1~s for the internal shock model), and $N_{\mbox{\small on}}$ is the
number of $\gamma$-ray bursts observed (1000 bursts will involve about
three years of observations with, e.g., SWIFT).  The advanced LIGO bound
described above corresponds to the conversion of
$\sim0.3\,\mbox{M}_\odot$ to gravitational waves at
$z\simeq1/2$.

\section*{Stochastic signals}

A stochastic gravitational wave signal, in a single detector, is
indistinguishable from detector noise. It is only when we have two or
more independent detectors that we can begin to distinguish between
detector noise and a stochastic signal: the stochastic signal, arising
as a superposition of plane waves from different directions, will
appear as correlated noise in pairs of detectors, with the correlation
rolling off for wavelengths greater than half the distance between the
detectors \cite{christensen92a,flanagan93a}. For the LIGO detector-pair
the correlation of the two detectors to a stochastic signal has a null
at approximately 64~Hz and a second null at approximately twice that
frequency. Beyond that second null there is very little power in the
correlated response of the detector pair to the incident stochastic
signal. 

It is convenient to characterize the strength of a stochastic
background by its energy density in a logarithmic frequency band,
relative to the closure density of the universe $\rho_0$:
\begin{equation}
\Omega_{\mbox{\small GW}} \equiv {f\over \rho_{0}}
  {d\rho_{\mbox{\small GW}}\over
  df},
\end{equation}
where $\rho_{\mbox{\small GW}}$ is the energy density in gravitational
waves. 

When considering sources of detectable stochastic gravitational
radiation it is conventional to consider signals of primordial origin:
e.g., radiation arising during an inflationary epoch or from
the decay of a cosmic string network (see \cite{maggiore00a} for an
excellent review).  Primordial nucleosynthesis places a bound on the
contribution to $\Omega_{\mbox{\small{}GW}}$ from these primordial sources of
$10^{-5}$: a larger $\Omega_{\mbox{\small{}GW}}$ would lead to a significantly
larger He$_4$ abundance than is observed. This bound is weak compared
to the theoretical predictions on the size of these backgrounds:
$\Omega_{\mbox{\small{}GW}}\simeq10^{-14}$ for inflation and
$\Omega_{\mbox{\small{}GW}}\lesssim10^{-11}$ for a cosmic string network. 

It is important to note, however, that \emph{the nucleosynthesis bound
  does not address gravitational waves produced after
  nucleosynthesis,} and there are a number of mechanisms capable of
producing significant stochastic gravitational waves at late times. As
important, these mechanisms are not exotic ones: they are as simple as
the confusion limit of discrete but unresolved conventional
astronomical sources of gravitational waves: e.g., core-collapse
supernovae, binaries or binary inspiral, or coalescence, etc..
Stochastic sources like these are certainly important for the LISA
detector\footnote{LISA will be overwhelmed at frequencies below
  $10^{-2.5}$~Hz by the radiation from unresolved close white dwarf
  binaries and other galactic binary systems \cite{hils90a}.}
\cite{folkner98a} and they may also be for the LIGO detector: it all
depends on the unknown event rate and luminosity of the individual
sources.

To assess the science reach of the LIGO detector pair to a stochastic
gravitational wave background, we ask what upper limit we can expect
to place on $\Omega_{\mbox{\small{}GW}}$ \emph{today}, based on the observed
correlation in the two LIGO detectors. Focusing on an observation
involving $1/3$~y of data and insisting on a 99\% confidence, we find
\cite{christensen92a,flanagan93a,allen99b,feldman98a} 
\begin{equation}
\Omega_{99\%} = \left\{
\begin{array}{lr}
3.3\times10^{-6}/h^2&\mbox{initial LIGO}\\
2.0\times10^{-9}/h^2&\mbox{advanced LIGO}
\end{array}
\right.
\end{equation}
where $h$ is the Hubble constant in units of 100~Km/s/Mpc. Thus, the
initial LIGO detectors can improve on the existing in-band limit on a
primordial background by a factor of approximately 3, and set a
completely new limit on a background from the confusion limit of more
conventional astronomical sources. The proposed advanced LIGO will be
able to improve on these bounds by a factor of just over 1000,
challenging some of the more optimistic proposals for a stochastic
background from conventional sources \cite{maggiore00a}. 

\section*{Core-collapse supernovae}

As a final example, consider the gravitational waves that arise from
the collapse of the stellar core in, e.g., a type II supernovae
explosion. The gravitational radiation signature of this source is
very uncertain: different plausible models for the progenitor give
very different gravitational radiation waveforms \cite{zwerger98a}. The spectrum of the
different bursts are more similar: the burst is roughly white to a
KHz, with the energy falling very rapidly at high frequencies.
Without assuming a waveform for the gravitational wave burst we can
compare the cross-correlation of the LIGO detector outputs at the
expected time of the gravitational wave burst with its value at other
times, not associated with a supernova. The difference of the actual
cross-correlation from its mean value can be interpreted, in this
model, as a measure of the mean-square $h$ averaged over the detectors
band. From the distance to the supernova (and, again, assuming that
the spectrum of $h$ is white to 1~KHz) we can convert this mean-square
amplitude into a measure of the fraction of the stellar cores rest
mass (assumed to be $\simeq1.4\,\mbox{M}_\odot$) converted into
gravitational waves. That efficiency $\epsilon$ we take to be the
``science reach'' of the detectors for radiation from core-collapse
supernovae. 

The limit we can place on $h^2$ in the band, and thus on $\epsilon$,
is better the better we know when the core collapse took place.  If
the core-collapse is galactic then we expect to be able to detect the
neutrinos directly, allowing us to fix the time of the core collapse
to under a second; on the other hand, if the supernova is
extragalactic, then we will have to rely on a backward extrapolation
of the light curve, which will allow us to fix the time of the
collapse to within an hour. For a supernova at a distance of 55~Kpc
(the large Magellanic cloud), the 95\% upper
limit we can place on $\epsilon$ is 
\begin{equation}
\epsilon_{95\%} = \left\{
\begin{array}{lr}
2\times10^{-4}&\mbox{initial LIGO}\\
2\times10^{-6}&\mbox{advanced LIGO}
\end{array}
\right.
\end{equation}

Galactic supernova occur at the rate of approximately 1 per 30~y. At
the distance of the center of the Virgo cluster --- 15~Mpc --- the
rate is on order 3~y${}^{-1}$. At this distance the advanced LIGO can
place an upper bound of $\epsilon_{95\%}\simeq24\%$ on the efficiency:
a physical upper bound (i.e., one less than 100\%).

Theoretical prejudice --- which, on this source, has revised itself by
several orders of magnitude many times over the past three decades ---
currently estimates the efficiency at $\epsilon<10^{-7}$--$10^{-8}$
\cite{finn90b,finn91a,monchmeyer91a}. 
These estimates are based on two-dimensional calculations and focus
only on the earliest parts of the collapse. As such they exclude the
radiation that might be associated with axisymmetries in the collapse
or in the rapid convective overturn of the hot proto-neutron star,
both of which may be significant sources of gravitational
radiation. Thus, while LIGO observations don't approach these
theoretical prejudices, they are nevertheless important for the
challenge they make to the prejudice. 

\section*{Summary}

I like to say that the initial LIGO detectors bound the
possible. In the case of binary inspiral they challenge some of the most
optimistic scenarios for the rate density of compact binary neutron
star and black hole systems; they set meaningful
upper limits on the crustal deformation of nearby, young neutron
stars; they improve the in-band limits on the strength of a
primordial stochastic gravitational wave  background and set a first
bound on the background strength owing to the confusion limit of more
conventional astronomical sources; and, in the event of a galactic
supernova while LIGO is ``on the air'', will set a physically
meaningful upper limit on the efficiency with which core collapse
supernovae produce gravitational waves. 

The proposed advanced LIGO challenges theory. It will survey a volume
of space that all but assures it should see several NS/NS binary
inspirals per year and also establish the rate density of NS/BH and
BH/BH binary systems. It will measure, or further improve the upper
limit on, the deformation of the crust of nearby neutron stars;
improve the limits on the stochastic background signal by three orders
of magnitude beyond initial LIGO;  may well explore the
$\gamma$-ray burst model; and place physical bounds on the supernova
efficiency. 

\bigskip

It is a pleasure to thank Joan Centrella for organizing an absolutely
splendid workshop, and for her seemingly boundless patience in waiting
for my contribution to this proceedings. The work described here was
supported by National Science Foundation awards PHY~98-00111 and
PHY~99-96213 and its predecessor PHY~95-03084.


\end{document}